\documentclass[twoside,11pt]{7ssmmp} 

\usepackage{amsfonts}
\usepackage{amssymb}
\usepackage{amsmath}
\usepackage{fancyhdr}    
\usepackage{graphicx}
\usepackage{rotating}

\def\g{ {\mathfrak g} } 
\def\G{ {\mathfrak G} } 
\def\F{ {\cal F} }
\def\bra#1{\left\langle #1\right|}
\def\ket#1{\left| #1\right\rangle}
\def\bracket#1#2{\left\langle #1 | #2 \right\rangle}

\newtheorem{theo}{Theorem}
\newtheorem{prop}{Proposition}
\newtheorem{cor}{Corollary}

\begin{document}
 \baselineskip=11pt

\title{Some representations of planar Galilean conformal algebra\hspace{.25mm}\thanks{\,Work supported by a grants-in-aid from JSPS (Contract No.23540154).}}
\author{\bf{N. Aizawa}\hspace{.25mm}\thanks{\,e-mail address:
aizawa@mi.s.osakafu-u.ac.jp}
\\ \normalsize{Department of Mathematics and Information Sciences, }\\
\normalsize{Graduate School of Science, Osaka Prefecture University, }\\
\normalsize{Nakamozu Campus, Sakai, Osaka 599-8531, Japan} 
}

\date{}

\maketitle

\begin{abstract}
 Representation theory of an infinite dimensional Galilean conformal algebra 
introduced by Martelli and Tachikawa is developed. We focus on the algebra 
defined in $(2+1)$ dimensional spacetime and consider central extension. 
It is then shown that the Verma modules are irreducible for non-vanishing 
highest weights. This is done by explicit computation of Kac determinant. 
We also present coadjoint representations of the Galilean conformal algebra and 
its Lie group. As an application of them, a coadjoint orbit of the Galilean conformal 
group is given in a simple case.

\end{abstract}

\section{Introduction}

 Representation theories of infinite dimensional Lie algebras such as Virasoro and Kac-Moody are, as is widely known, 
fruitful field of mathematical and physical research. Those Lie algebras are associated with certain semisimple 
finite dimensional Lie algebras. 
Recent renewed interest in conformal algebras in nor-relativistic setting, especially 
in the context of non-relativistic AdS/CFT correspondence \cite{NRAdS,MT}, introduced a new class of 
infinite dimensional Lie algebras associated with certain non-semisimple Lie algebras \cite{MT,Hen,InfGal}.  
The best studied example of this class of algebra is the Schr\"odinger-Virasoro algebra \cite{Hen} 
associated with six dimensional Schr\"odinger algebra. 
Readers may refer the book \cite{SchVir} for background, definition and representation theory of 
Schr\"odinger-Virasoro algebra. 

 In the present work we investigate representations of an another algebra of this class 
introduced in \cite{MT}. 
In \cite{MT} a two-parameter family of infinite dimensional algebra is 
discussed. One of the parameters is the dimension of space $d$ and the other is ``spin'' $ \ell. $ 
Here we pick up the algebra corresponding to $ d = 2, \ \ell = 1 $ and denote it by $ \g_0. $
The basis of $ \g_0 $ is $ L_m, P_n^i $ and $ J_m $ with 
$ m \in {\mathbb Z}, \ i, j = 1, 2. $ The defining relations of the algebra are given by
\begin{eqnarray}
  & & [L_m, L_n] = (m-n) L_{m+n},
    \nonumber \\
  & & [L_m, P_n^i] = (m-n) P_{m+n}^i,   \qquad [L_m, J_n] = -n J_{m+n},
    \label{alg_def}\\
  & &  [J_m, P_n^i] = \sum_j \epsilon_{ij} P_{m+n}^j, \qquad\quad  [J_m, J_n] =  [P_m^i, P_n^j] = 0,
  \nonumber
\end{eqnarray}
where $  \epsilon_{12} = -\epsilon_{21} = 1. $ 
It is seen that $ \langle L_m \rangle $ form a centerless Virasoro algebra and $ \langle P_n^i \rangle, \langle J_n \rangle  $
carry a representation of it.  Moreover, $ \langle P_n^i \rangle $ is an Abelian ideal of $ \g_0. $ 
Following vector field realization of $ \g_0 $ is helpful to see the algebra is indeed related to $(2+1)$ dimensional 
spacetime:
\begin{eqnarray}
  & & L_m = -t^{m+1} \partial_t -  (m+1) t^m  x_i \partial_{x_i},  \nonumber
  \\
  & & J_m = -t^m (x_1 \partial_{x_2} - x_2 \partial_{x_1}), \qquad
      P^i_m = - t^{m+1} \partial_{x_i}.  \label{Eq:inf-vec}
\end{eqnarray}

  The subspace spanned by $ \langle L_0, L_{\pm 1}, J_0, P^i_0, P^i_{\pm 1} \rangle $ forms a ten dimensional subalgebra also called 
Galilean conformal algebra. 
It is known that this subalgebra has a peculiar central extension in the sense that it exists for only $ d = 2 $ and 
integral values of $ \ell $ \cite{MT,exotic}. This {\it exotic} central extension makes the Abelian ideal $ \langle P_m^i \rangle $ 
makes non-Abelian:
\[
  [P^i_m, P^j_n] = I_{mn} \epsilon^{ij} \Theta,
\]
where $ I_{mn} $ is a symmetric tensor. 
It is also known that the exotic Galilean conformal algebra has some physical applications \cite{MT,exotic}. 
(For more details on finite dimensional Galilean conformal algebras, see \cite{CGA} and references therein). 
In fact, these observation on the finite dimensional exotic Galilean conformal algebra is a motivation of the 
present work. The algebra $ \g_0 $ is an infinite dimensional version of the 
Galilean conformal algebra for $ d = 2, \ \ell = 1. $ 
Then natural questions arise. Dose $ \g_0 $ have a central extension of exotic type ? 
In what kind of physical context $ \g_0 $ appears ?  
We shall give a negative answer to the first question: $ \g_0 $ does not have the exotic central extension 
but others. We have no idea on the second question at this writing. 
However, we know that physical application of algebraic object is available via its representations. 
Therefore we study representations of $ \g_0 $ with central extensions.  

  The plan of this paper is as follows: In the next section, we introduce central extensions of $ \g_0 $ 
and denote the algebra with the central extensions by $ \g. $ We then consider highest weight representations 
of $ \g, $ especially we investigate irreducibility of Verma modules over $ \g. $ 
This is done by calculating Kac determinant. We give a explicit formula of the Kac determinant and 
it shows that the Verma modules are irreducible for non-vanishing highest weights. 
In \S 3 we investigate coadjoint representations of $\g. $ By employing the {\it regular dual} of $ \g $ as 
a space $\g^*$ dual to $ \g, $ we calculate the action of $ \g $ on $ \g^*. $ 
This allows us to compute coadjoint orbit of the infinite dimensional group $ \G $ having $ \g $ as its 
Lie algebra. We shall give a simple example of coadjoint orbit. 
However, the coadjoint action of $ \g $ is not enough to classify all coadjoint orbits of $ \G $ \cite{SchVir}. 
We need coadjoint representation of the group $ \G. $ This is presented in \S 4.   

%
\section{Verma modules and its irreducibility}

  We start with central extensions of $ \g_0. $ 
Our interest is in the possibility of the exotic central extension which makes Abelian 
ideal non-Abelian. It is also curious whether the Virasoro subalgebra has the ordinary 
central extension. We here present only the result and the proof is found in \cite{NA}.
\begin{prop}
 The algebra $\g_0 $ has the following central extensions
 \begin{eqnarray}
    & &  [L_m, L_n] = (m-n) L_{m+n} + \frac{\alpha}{12} m(m^2-1) \delta_{m+n,0},  \\
    & &  [J_m, J_n] = \beta m \delta_{m+n,0}, 
 \end{eqnarray}
where $ \alpha,\beta $ are central charges. However, the exotic type central extension 
is impossible. 
\end{prop}
We denote the algebra with the central extensions by $ \g. $ 
In the rest of this paper we investigate representation of  $ \g. $

  Our main subject in this section is highest weight representations of $ \g, $ 
especially we consider Verma modules. 
Important fact is that the algebra $ \g $ admits the triangular decomposition. 
Define the degree of  $X_n \in \g $ by $ \deg(X_n) = -n $ where $ X = L, J, P^i. $ 
This allows us to define the triangular decomposition of $ \g: $ 
\begin{eqnarray}
 \g &=& \g^- \oplus \g^0 \oplus \g^+
    \nonumber \\
    &=& 
 \langle\; L_{-n}, J_{-n}, P_{-n}^i \; \rangle  \oplus  \langle\; L_{0}, J_{0}, P_{0}^i \; \rangle  
     \oplus  \langle\; L_{n}, J_{n}, P_{n}^i \; \rangle, \quad 
     n \in {\mathbb N}
    \label{tri-decomp}
\end{eqnarray}
Let $ \ket{0} $ be the highest weight vector:
\begin{equation}
    \g^+ \ket{0} = 0, \quad L_0 \ket{0} = h \ket{0}, \quad 
    J_0 \ket{0} = \mu \ket{0}, \quad P_0^i \ket{0} = \rho_i \ket{0}.
    \label{HWvector-def}
\end{equation}
Following the usual definition of Verma modules, we define the Verma modules over $ \g $ by
\begin{equation}
 V^{\chi} = U(\g^-) \otimes \ket{0}, \qquad \chi = (h, \mu, \rho_1, \rho_2, \alpha, \beta),
 \label{VMdef}
\end{equation}
where $ U(\g^-) $ denotes the universal enveloping algebra of $ \g^-. $ 
The Verma module $ V^{\chi} $ is a graded-modules through a natural 
extension of the degree from $ \g $ to $ U(\g) $ by 
$ \deg(XY) = \deg(X) + \deg(Y),$ $ \; X, Y \in U(\g), $
\[
   V^{\chi}
       = \bigoplus_{ n \in {\mathbb Z}_{\geq 0} } V^{\chi}_n,
     \quad
     V^{\chi}_n = \{ X \ket{0} \ | \ X \in U(\g^-), \ \deg(X) = n \ \}.
\]
There exists an algebraic anti-automorphism $ \omega : \g \to \g $ defined by
\begin{equation}
   \omega(L_m) = L_{-m}, \qquad \omega(J_m) = J_{-m}, \qquad \omega(P_m^i) = P_{-m}^i.
   \label{Eq:antiauto}
\end{equation}
One can introduce an inner product in $ V^{\chi} $ by extending the anti-automorphism 
$ \omega $  to $ U(\g). $ 
We define the inner product of $ X \ket{0}, Y \ket{0} \in V^{\chi} $ by
\[
   \bra{0} \omega(X) Y \ket{0}, \qquad \bracket{0}{0} = 1. 
\]

 The reducibility of  $ V^{\chi} $ may be investigated by the Kac determinant. 
The Kac determinant is defined as usual~\cite{KaRa}. 
Let $ \ket{i} (i = 1, \cdots, \mbox{dim} V_n^{\chi}) $ be a basis of $ V_n^{\chi}, $ then 
the Kac determinant at level (degree) $n$ is given by
\[
  \Delta_n = \det (\; \bracket{i}{j} \;).
\]
We want to calculate $ \Delta_n.$ 
The main obstacles of this calculation are rapid increase of 
dim $V^{\chi}_n $ as a function of $n$ and $ \Delta_n$ is never 
reduced to the determinant of diagonal matrix. 
For illustration we list $ \mbox{dim}V^{\chi}_n $  for some small $n$: 
\[
  \begin{array}{c|cccccc}
    n & \quad 0 & \quad 1 &\quad 2 &\quad 3 &\quad 4 & \quad 5\\ \hline
    \mbox{dim}V^{\chi}_n\quad & \quad 1 &\quad 4 &\quad 14 &\quad 40 &\quad 105 & \quad 252
  \end{array}
\]
However, one can carry out the computation by the method similar to that for
Schr\"odinger-Virasoro algebra used in \cite{SchVir}. We here merely mention the result and 
do not go into the detail. 
The computational details are found in \cite{NA}. 
To mention the result we need some preparation. 

A partition $ A=(a_1a_2\cdots a_{\ell}) $ of a positive integer $a$ is the sequence of 
positive integers such that
\begin{eqnarray*}
  a &=& a_1 + a_2 + \cdots + a_{\ell},  \\
    & & a_1 \geq a_2 \geq \cdots \geq a_{\ell} > 0.  
\end{eqnarray*}
The integers  $\ell$ is called length of the partition $A$ and denoted by  $ \ell(A). $ 
For a given partition $A$ of $a$, we decompose a set of integers $ a_1, a_2, \cdots, a_{\ell} $ to two subsets 
by selecting $s $ integers from them $ ( 0 \leq s \leq \ell)$:
\begin{equation}
  {A}_1 = \{\ a_{\sigma_1} \geq a_{\sigma_2} \geq \cdots \geq a_{\sigma_s} \ \}, \qquad
  {A}_2 = \{\ a_{\rho_1} \geq a_{\rho_2} \geq \cdots \geq a_{\rho_{\ell-s}} \ \}. 
  \label{Eq:subpar}
\end{equation}
$ A_1 $ consists of the selected $ s $ integers and the members of $ A_2 $ are the rest integers 
so that the partition $ A $ is decomposed into a pair of partitions $(A_1 A_2). $ 
We denote the number of all possible pairs $(A_1 A_2)$ by $ s(A). $ 
For instance, let $ A = (21) $ then the possible pairs are 
$ (A_1 A_2) =  ((21) \phi), \; ((2) (1)),\ ((1) (2)), \; (\phi (21)) $ so that 
$ s((21)) = 4. $ 
Now we mention our result of $ \Delta_n:$
\begin{theo}
\label{theo:KacDet}
 Level $n$ Kac determinant is given by
\[
  \Delta_n = c_n \prod_{a,b} \prod_{A,B} (\rho_1^2 + \rho_2^2)^{\frac{1}{2} s(A) s(B)( \ell(A) + \ell(B))}
\]
where the pair $ (a,b) $ runs all possible  non-negative integers satisfying $ n = a + b $ and 
the pair $ (A,B) $ runs all possible partitions of fixed $ a $ and $b.$   
The coefficient $ c_n $ is a numerical constant. 
\end{theo}
Explicit values of $ c_n $ (up to sign) and the power of $ \rho_1^2 + \rho_2^2 $ for 
$ n = 1, 2, 3 $ are listed below:
\[
  \begin{array}{cccc}
    n & 1 & 2 & 3 \\ 
    (c_n,\mbox{power})\ &\ (2,2)\ &\ (2^{18},12)\ &\ (2^{72} 3^6, 48)
  \end{array}
\]
We remark that 
$ \Delta_n $ is independent of the central charges $ \alpha, \beta. $ 
Thus the formula of $ \Delta_n $ is common for the algebras $ \g $ and $ \g_0. $ 
We see from Theorem 1 that if $ \rho_1^2 + \rho_2^2 \neq 0 $ 
then the Kac determinant $ \Delta_n $ never vanish so that 
there exist no singular vectors in $ V_n^{\chi} $ for any $n.$
\begin{cor}
 The Verma module $ V^{\chi} $ is irreducible if $ \rho_1^2 + \rho_2^2 \neq 0. $ 
\end{cor}
This is a sharp contrast to the corresponding finite dimensional algebra with the exotic central 
extension where some Verma modules for certain non-vanishing highest weights are reducible~\cite{Phil}. 
However, similar irreducibility of Verma modules for nonvanishing highest weights is also observed for 
the Schr\"odinger-Virasoro algebra \cite{SchVir}.

%
\section{Coadjoint representation of $\g$ and coadjoint orbit}

  Our next subject is coadjoint representations. 
Physical implication of coadjoint representations is laid in coadjoint orbits of 
the infinite dimensional Lie {\it group} $ \G $ which is an integration of $\g. $
As is known widely, any coadjoint orbit is  a symplectic manifold with Poisson structure 
so that the base of geometric quantization \cite{Kir}. 
We thus want to derive coadjoint representation of $ \G $ and make a classification of 
coadjoint orbits. However, this work is in progress. 
Here we shall give a simple example of the coadjoint orbit using a coadjoint representation 
of $ \g $ which is calculated by employing the method in \cite{SchVir}.
  
Coadjoint representations are, by definition, obtained by the action of $ \g $ or $ \G $ on 
the space $ \g^* $ dual to $ \g. $ 
Thus we have to determine the space $ \g^*. $ 
To this end, we return to the centerless algebra $ \g_0$ and recast it in the form of the current:
\[
      L_f = f(\theta) \partial_{\theta} + f'(\theta) x_i \partial_{x_i}, \quad 
      J_f = f(\theta) (x_1 \partial_{x_2} - x_2 \partial_{x_1}), \quad P^i_f = f(\theta) \partial_{x_i},
\]
where $ \theta \in S^1 $ is a compactified time coordinate. 
Then we have a current algebra with the commutation relations:
\begin{eqnarray}
   & & 
   [L_f, L_g] = L_{fg'-f'g}, \quad   [L_f, P_g^i] = P_{fg'-f'g}^i, \quad    [L_f, J_g] = J_{fg'}, 
   \nonumber \\
   & & \quad\ \ 
    [J_f, P_g^i] = -\sum_{k} \epsilon_{ik} P^k_{fg}, \quad
    [J_f, J_g] = [P_f^i, P_g^j] = 0.
   \label{CAcom}
\end{eqnarray}
By considering Fourier components one may recover the relations (\ref{alg_def}) and (\ref{Eq:inf-vec}). 
Now let us recall that the centerless Virasoro algebra has the one-parameter family of 
representation on the space of {\it densities} $ \F_{\lambda} $ of the form $ \phi(\theta) d\theta^{-\lambda} $ \cite{SchVir} 
(see also \cite{KaRa}). The action of $ L_f$ on $ \F_{\lambda} $ is defined by
\begin{equation}
 L_f (\; \phi(\theta) d\theta^{-\lambda} \;) := (f \phi' - \lambda f' \phi) d\theta^{-\lambda}.
 \label{1param-rep}
\end{equation}
The dual space $ {\cal F}_{\lambda}^* $ may be identified with ${\cal F}_{-1-\lambda} $ through the paring: 
\begin{eqnarray}
  & & \F_{\lambda}^* \times \F_{\lambda} \to {\mathbb C}, \nonumber \\
  & & 
  \langle \; u(\theta) d\theta^{1+\lambda}, \; f(\theta) \theta d\theta^{-\lambda} \; \rangle = \int_{S^1} u(\theta) f(\theta) d\theta.
  \label{Eq:paring}
\end{eqnarray}
By comparing (\ref{CAcom}) and (\ref{1param-rep}) we are allow to make the following identification:
\[
 L_g \simeq \F_1, \quad P^i_g \simeq \F_1, \quad J_g \simeq \F_0. 
\]
Thus $ \g_0 \simeq \F_1 \oplus \F_1 \oplus \F_1 \oplus \F_0 $ as a vector space. 
It follows that
\[
    \g_0^* \simeq \F_{-2} \oplus \F_{-2} \oplus \F_{-2} \oplus \F_{-1}
\]

Next we consider the central extensions in Proposition 1. 
The algebra $ \g $ may be regarded as 
\[
    \g \simeq \g_0 \oplus {\mathbb R} \oplus {\mathbb R}.
\]
Therefore, we obtain the algebra dual to $ \g $ as follows:
\begin{equation}
   \g^*  \simeq \F_{-2} \oplus \F_{-2} \oplus \F_{-2} \oplus \F_{-1} \oplus {\mathbb R} \oplus {\mathbb R}.
  \label{dualspace}
\end{equation}
We denote an element of $ \g^* $ by
\begin{equation}
    \gamma_0 d\theta^2 + \gamma_1 d\theta^2 + \gamma_2 d\theta^2 + \gamma_3 d\theta + a + b \ \in \ \g^*
\end{equation}
and identify it with the column vector 
$ \vec{\gamma} = {}^t(\gamma_0, \gamma_1, \gamma_2, \gamma_3, a,  b). $
Then the duality pairing is given by
\begin{equation}
    \langle \vec{\gamma}, \vec{X} \rangle = \sum_{i=0}^3 \int_{S^1} \gamma_i f_i\, d\theta + a \alpha + b \beta,
    \label{Eq:pairing}
\end{equation}
where $   \vec{X} = ( L_{f_0}, P_{f_1}^1, P_{f_2}^2, J_{f_3}, \alpha,  \beta). $

  We now come to the definition of coadjoint action of $ \g $ on $ \g^*. $ 
Define an action $ X(\gamma) $ of $ \g $ on $ \g^* $ by
\begin{equation}
       \langle X(\gamma), Y \rangle := -\langle \gamma, [X,Y] \rangle, \quad
       X, Y \in {\mathfrak g}, \ \gamma \in {\mathfrak g}^*,
       \label{coadjaction-def}
\end{equation}
then $ X(\gamma) $ gives a representation of $ \g $ \cite{Kir}. 
It is not difficult to calculate the coadjoint action according to the definition. 
The result is summarized as follows.
\begin{prop}
Coadjoint representation of $ \g $ is given by
\begin{eqnarray}
 & & L_{f_0}
    \begin{pmatrix}
      \gamma_0 \\ \gamma_1 \\ \gamma_2 \\ \gamma_3
    \end{pmatrix}
     = 
    \begin{pmatrix}
      a f_0''' + 2 \gamma_0 f_0' + \gamma_0' f_0 \\
      2 \gamma_1 f_0' + \gamma_1' f_0 \\
      2 \gamma_2 f_0' + \gamma_2' f_0 \\
      \gamma_3 f_0' + \gamma_3' f_0
    \end{pmatrix},
  \quad
  J_{f_3}(\vec{\gamma})  = 
   \begin{pmatrix}
       \gamma_3 f_3' \\
       \gamma_2 f_3 \\
       -\gamma_1 f_3 \\
       bf_3'
   \end{pmatrix},
 \nonumber \\
 & & P_{f_1}^1(\vec{\gamma}) = 
   \begin{pmatrix}
      2 \gamma_1 f_1' + \gamma_1' f_1 \\
      0 \\ 0 \\
      -\gamma_2 f_1
   \end{pmatrix},
   \qquad
   P_{f_2}^2(\vec{\gamma}) = 
   \begin{pmatrix}
      2 \gamma_2 f_2' + \gamma_2' f_2 \\
      0 \\ 0 \\
      -\gamma_1 f_2
   \end{pmatrix}.
   \label{coadrep-g}
\end{eqnarray}
The coadjoint action on the central elements  $ a, b $ is trivial 
so that they are omitted. 
\end{prop}
We remark that $ L_{f_0}(\gamma_0) $ is identical to the coadjoint action of 
the Virasoro algebra \cite{Witten,Ba}. 

 Using the formulae in Proposition 2 one may calculate coadjoint orbits. By definition, 
coadjoint orbit is an orbit of the group $ \G $ in the space $ \g^*. $ 
To find the orbits we have to find an isotropy group $ {\mathfrak H} \subset \G $ for a given $ \vec{\gamma} \in \g^*. $ 
Then one may identify the orbit with $ \G/{\mathfrak H}. $ 
Here we consider an infinitesimal transformation of $ \vec{\gamma} $ by $ \G, $ namely, we look for an 
isotropy algebra of $ \vec{\gamma}. $ 
To this end, one has to solve the equation:
\begin{equation}
   (L_{f_0}+ J_{f_3}- P_{f_1}^1 - P_{f_2}^2)(\vec{\gamma}) = 0. 
  \label{iso-alg}
\end{equation}
We show only a simple example of the solution of the equation (\ref{iso-alg}). Full classification of 
the coadjoint orbits of the group $ \G $  will be presented elsewhere. 
We assume that $ \gamma_1 $ and $ \gamma_2 $ are nonvanishing constants. 
Then the equation (\ref{iso-alg}) is reduced to a set of four equations. 
\begin{eqnarray}
  & & a f_0''' + 2 \gamma_0 f_0' + \gamma_0' f_0 + \gamma_3 f_3' = 2( \gamma_1 f_1' + \gamma_2 f_2'),
   \nonumber \\
  & & 2 \gamma_1 f_0' + \gamma_2 f_3 = 0,
   \nonumber \\
  & & 2 \gamma_2 f_0' - \gamma_1 f_3 = 0,
   \nonumber \\
  & & \gamma_3 f_0' + \gamma_3' f_0 + b f_3' = -\gamma_2 f_1 + \gamma_1 f_2.
   \nonumber
\end{eqnarray}
This set of equation is easily solved to give
\begin{equation}
     f_0 = const.,\quad f_3 = 0, \quad
     f_k = \Phi_k(\theta) f_0 + const, \ k = 1, 2
\end{equation}
where $ \Phi_k(\theta) $ is a functions of $ \gamma_0(\theta), \ \gamma_3(\theta). $ 
We thus have the following generators of the isotropy algebra:
\begin{equation}
 L_{f_0} = const. \partial_{\theta}, \qquad P_{f_k}^k = f_k(\theta) \partial_{x_k}, \qquad
    J_{f_3} = 0.
    \label{gen-iso-alg}
\end{equation}
One may see that $ L_{f_0} $ generate a translation on $ S^1 $ and $ f_k(\theta) $ with $ k = 1, 2 $ is an  
arbitrary function.  It follows that the isotropy group for $ \gamma_1, \gamma_2 $ being nonvanishing constant 
is a semidirect product of $ U(1) $ by the group $ {\mathfrak G}_P $ generated by $ P^k_{f_k}. $ 
Recalling that the group $ \G $ is a semidirect product of the Virasoro group Diff$(S^1)$ by $ {\mathfrak G}_P $ and 
$ \widehat{SO(2)}, $ we conclude that the coadjoint orbit for $ \gamma_1, \gamma_2 $ being non-vanishing constant is 
a product of a generic orbit Diff$(S^1)$ of the Virasoro group by $ \widehat{SO(2)}. $ 

%
\section{Coadjoint representation of $ \G $}

In this section we give the group $ \G  $ more explicitly and 
calculate its coadjoint action which will be used to 
determine all coadjoint orbits of $\G. $ 
The Lie algebra $ \g $ is a semidirect sum of the Virasoro algebra and 
the algebra $ \langle P_{f_k}^k, J_{f_3} \rangle. $ 
Since integration of the Virasoro algebra is well-known, 
we first consider the group generated by $ P_{\eta_k}^k $ and $  J_{\xi}. $ 
We denote an element of the group by $ ( \xi, \eta_1, \eta_2). $
Regarding  $ \exp( J_{\xi} ) \exp( P_{\eta_1}^1 + P_{\eta_2}^2) $ as 
$ ( \xi, \eta_1, \eta_2), $ 
one may deduce the following formulae for group multiplication and inverse:
\begin{eqnarray}
       & & 
        (\xi,\eta_1,\eta_2) (\rho, \sigma_1, \sigma_2) = 
        \nonumber \\
       & & \qquad 
        (\xi+\rho, \sigma_1 + \eta_1 \cos \rho - \eta_2 \sin \rho, \sigma_2 + \eta_1 \sin\rho + \eta_2 \cos\rho) \exp( c \beta),
        \nonumber \\
       & &  (\xi,\eta_1,\eta_2)^{-1} = (-\xi,-\eta_1 \cos\xi - \eta_2 \sin\xi, \eta_1 \sin\xi - \eta_2 \cos\xi). 
        \nonumber
\end{eqnarray}
The group $ \G  $ is defined by
\begin{equation}
 (\phi,\xi,\eta_1, \eta_2) := (id, \xi,\eta_1, \eta_2) (\phi,0,0,0), \qquad \phi \in \mbox{Diff}(S^1)
  \label{G-def}
\end{equation}
It is not verify the relation
\begin{equation}
  (\phi,0,0,0) (id, \xi,\eta_1, \eta_2) = 
  \left(  \phi, \xi \circ \phi, \frac{ \eta_1 \circ \phi}{\phi'}, \frac{ \eta_2 \circ \phi}{\phi'} \right).
  \label{G-def2}
\end{equation}

  Coadjoint action of $ \G $ on $ \g^* $ is defined by \cite{Kir}
\begin{equation}
       \langle g(\gamma), X \rangle := \langle \gamma, g^{-1} X g \rangle, \quad 
       X \in \g, \quad \gamma \in \g^*, \quad g \in \G
     \label{G-action}
\end{equation}
This gives a representation of $ \G. $ 
Now we are ready to calculate coadjoint action of $ \G. $ We treat the Virasoro subgroup separately.
\begin{prop}
Coadjoint action of {\rm Diff}$(S^1)$ on $ \g^* $ is given by
\[
  \phi(\vec{\gamma}) = 
  \left(
     \begin{array}{c}
        a \Theta(\phi) + (\gamma_0 \circ \phi) (\phi')^2 \\[3pt]
         (\gamma_1 \circ \phi) (\phi')^2 \\[3pt]
         (\gamma_2 \circ \phi) (\phi')^2 \\[3pt]
         (\gamma_3 \circ \phi) \phi' 
     \end{array}
  \right),
  \quad 
  \Theta(\phi) = \frac{ \phi''' }{\phi'} - \frac{3}{2} \left(  \frac{\phi''}{\phi'}  \right)^2,
\]
where $ \Theta(\phi) $ denotes the Schwarzian derivative of $ \theta. $ 
Action on the central elements is omitted.
\end{prop}
One may easily prove Proposition 2. $ \phi(\gamma_0) $ is nothing but the well-known coadjoint action 
of the Virasoro group \cite{Witten,Ba}. Denoting the coordinate change on $ S^1 $ due to  Diff$(S^1)$  by  
$ \theta = \phi(t), $ we have
\[
 \gamma(\theta) d\theta^{\lambda} = \gamma(\phi(t)) \left( \frac{d\theta}{dt} dt \right)^{\lambda} = (\gamma \circ \phi) (\phi')^{\lambda} 
   dt^{\lambda}.
\]
$ \phi(\gamma_k) $ for $ k = 1, 2, 3 $ is immediately obtained from this relation. 

 Let us turn to coadjoint action of the $(\xi,\eta_1,\eta_2)$-subgroup. 
It is calculated straightforwardly by using definition (\ref{G-action}). 
As an illustration we compute $ (\xi,\eta_1,\eta_2)(\gamma_1). $ 
Let $ g = (\xi,\eta_1,\eta_2), $ then it is easy to see that 
\[
 g^{-1} P_{f_1}^1 g = P^1_{f_1} \; \cos\xi + P^2_{f_2}\; \sin\xi. 
\]
By (\ref{G-action})
\[
 \langle  g(\vec{\gamma}), P^1_{f_1} \rangle = 
  \langle \vec{\gamma}, g^{-1}  P^1_{f_1} g \rangle = 
  \int_{S^1} ( \gamma_1 \cos\xi + \gamma_2 \sin\xi) f_1 d\theta. 
\]
It follows that
$
  g(\gamma_1) = \gamma_1 \cos\xi + \gamma_2 \sin\xi.
$
$ g(\gamma_k) $ for $ k = 0, 2, 3 $ is computed in a similar way and 
result is summarized as follows:
\begin{prop}
Coadjoint representation of $(\xi,\eta_1,\eta_2)$-subgroup is given by
\begin{eqnarray*}
  & & (\xi,\eta_1,\eta_2)(\vec{\gamma}) = 
  \\
  & & \qquad 
  \left(
     \begin{array}{c}
        \gamma_0 + \sum_{k=1,2} (\gamma_k' \eta_k + 2 \gamma_k \eta_k') + (\gamma_1\eta_2-\gamma_2\eta_1 + \gamma_3) \xi' + \frac{b}{2}(\xi')^3
        \\[3pt]
        \gamma_1 \cos \xi + \gamma_2 \sin \xi 
        \\[3pt]
        -\gamma_1 \sin \xi + \gamma_2 \cos \xi
        \\[3pt]
        \gamma_1 \beta_2 - \gamma_2 \beta_1 + \gamma_3 + b \xi'
     \end{array}
  \right).
\end{eqnarray*}
Action on the central elements is omitted.
\end{prop}

%
\section{Concluding remarks}

  We studied  infinite dimensional Galilean conformal 
algebra introduced by Martelli and Tachikawa. 
We concentrated on the algebra defined in $(2+1)$ dimensional 
spacetime and  took into account the central extensions. 
Our main results are the proof of irreducibility of the Verma modules 
over $ \g $ and explicit formula of the coadjoint representations of 
$ \g  $ and $ \G. $ 
We also presented a coadjoint orbit of $ \G $ for a simple case. 

  The present work will be followed by a full classification of 
coadjoint orbit of $ \G. $ This will be done by using Proposition 3 and 4. 
Another interesting problem in  representation theory is action of $ \G $ on 
a space of differential operators. It is known that Virasoro group acts on 
the space of Hill operator and Hill operator gives a Lax form of KdV equation \cite{Ba}. 
It is also known that Schr\"odinger-Virasoro algebra acts on the space of 
Schr\"odinger operator \cite{SchVir}. Is it possible to represent $ \G $ on 
certain space of differential operator ? If so, what is the space ? 
Answer to this question would open a way to the most interesting problem on $ \g $ or $ \G, $ 
namely, physical applications. 

  We focus on the algebra in $(2+1)$ dimensional spacetime, however, algebra $ \g_0 $ is 
also defined for other dimensional spacetime. Furthermore, as mentioned in \S 1, 
there are some variety of infinite dimensional Galilean conformal algebra \cite{InfGal}. 
Nevertheless, only limited number of works has been published on those algebraic structure. 
They are waiting for extensive study from physical and mathematical points of view.


\end{document}